\documentclass{article}
\usepackage{amsmath}
\usepackage{float}
\usepackage[top=0.75in, bottom=0.75in, left=1.0in, right=1.0in]{geometry}

\begin{document}

\title{Temperature cooling in quantum dissipation channel and the
correspondimg thermal vacuum state\thanks{%
Work supported by the National Natural Science Foundation of China under
grant (No: 61203061, 61403362, 61374091, 61473199 and 11175113) }}
\author{Wang Yao-xiong$^{1}$, Gao Fang$^{1}$, Fan Hong-yi$^{1,2}$, and Tang
Xu-bing$^{1,3\dag }$ \\
%EndAName
$^{1}$Institute of Intelligent Machines, \\ Chinese Academy of Sciences,
Hefei 230031, China\\
$^{2}$Department of Material Science \& Engineering, \\ University of Science \
\  \& Technology of China, Hefei 230027, China\\
$^{3}$School of Mathematics \& Physics Science and Engineering, \\
Anhui University of Technology, Ma'anshan 243032, China\\
$^{\dag }$ttxxbb@ahut.edu.cn}
\maketitle

\begin{abstract}
We examine temperature cooling of optical chaotic light in a quantum
dissipation channel with the damping parameter $\kappa .$The way we do it is
by introducing its thermal vacuum state which can expose entangling effect
between the system and the reservoir. The temperature cooling formula is
derived, which depends on the parameter $\kappa ,$ by adjusting $\kappa $
one can control temperature.
\end{abstract}

\section{Introduction}

In nature most systems are immersed in their enviroments, energy exchange
between system and its enviroment always happens, this brings system's
dissipation with quantum decoherence \cite%
{Gardiner_2000_qn,Ankerhold_2003_lnp}. If these systems involve
non-negligible correlations amongst their components, quantum memory
(non-Markovian) effects cannot be ignored. If the feedback from enviroment
is extremely weak, we can say this process is Markovian, and Its dynamics
described by the master equation or the associated Langevin or Fokker-Planck
equations \cite{Scully_1998_qo,Orszag_2000_qo}. The Lindblad equation is the
most general form for a Markovian master equation, and it is very important
for the treatment of irreversible and non-unitary processes, from
dissipation \cite{Breuer_2002_oqs} and decoherence \cite{fan_2008_mplb} to
the quantum measurement process \cite{Mandel_1995_ocqo,Gardiner_2000_qn}.

In quantum optics and quantum statistics theory, a damping harmonic
osciallator in thermal bath is one of the most famous dissipation model. Its
dynamics of this model can be described by the Lindblad master equation. The
associated amplitude damping mechanism of system in physical processes is
governed by the following master equation \cite%
{Scully_1998_qo,Orszag_2000_qo}
\begin{equation}
\frac{d\rho \left( t\right) }{dt}=\kappa \left( 2a\rho a^{\dagger
}-a^{\dagger }a\rho -\rho a^{\dagger }a\right) ,  \label{1}
\end{equation}%
where $\rho $ is the density operator of the system, $\kappa $ is the rate
of decay, $a,a^{\dagger }$ are boson annihilation and creation operator,
respectively, $\left[ a,a^{\dagger }\right] =1$. Such an equation can be
conveniently solved by virtue of the entangled state representation \cite%
{fan_1994_pra}, and the solution is in so-called Kraus form
\begin{equation}
\  \rho \left( t\right) =\sum_{n=0}^{\infty }\frac{V^{n}}{n!}e^{-\kappa
ta^{\dagger }a}a^{n}\rho _{0}a^{\dag n}e^{-\kappa ta^{\dagger }a},  \label{2}
\end{equation}%
where $\rho _{0}$ is system's initial density operator,
\begin{equation}
V\equiv 1-e^{-2\kappa t}.  \label{3}
\end{equation}

As a common knowledge, during the thermal communication between system and
reservoir, excitation and de-excitation processes are influenced by the
exchange of energy between them. The loss of energy by the system can occur
in two ways: 1). It emits quanta with positive energy $\hbar \omega $
(denoted by an annihilation operator $a$). 2). By creating holes of
particles with positive energy in the reservoir. When the latter process
takes place, we say that a hole is created in the reservoir by $\tilde{a}%
^{\dagger }.$ The tilde mode $\tilde{a}$ (reservoir mode) is independent of
the system's mode $a,$ $\left[ \tilde{a},\tilde{a}^{\dagger }\right] =1.$

An interesting question that has been overlooked for long is how does the
temperature of system change when the system undergoes dissipation? To be
concrete, we consider when an optical chaotic field \cite{fan_2011_cpl},
described by the density operator%
\begin{equation}
\rho _{c}=\left( 1-e^{-\frac{\hbar \omega }{kT}}\right) e^{-\frac{\hbar
\omega }{kT}a^{\dagger }a},  \label{4}
\end{equation}%
here $\beta =\frac{1}{kT}$ $\left( k\text{ is the Boltzmann constant, }T%
\text{ is temperature}\right) $, undergoes amplitude damping described by
Eq. (\ref{1}), then how is its temperature evolves with time?

In order to answer this question we recall the Thermal Field Dynamics (TFD)
theory of Takahashi and Umezawa \cite{Takahashi_1996_ijmpb} for converting
the statistical average Tr$\left( A\rho \right) $ at nonzero temperature $T$
into equivalent pure state expectation value they introduced the state in a
doubled freedom Fock space
\begin{equation}
\sec \text{h}\theta \exp \left[ a^{\dagger }\tilde{a}^{\dagger }\tanh \theta %
\right] \left \vert 0\tilde{0}\right \rangle \equiv \left \vert 0(\beta
)\right \rangle  \label{6}
\end{equation}%
such that

\begin{equation}
\left \langle 0(\beta )\right \vert A\left \vert 0(\beta )\right \rangle =%
\text{Tr}\left( A\rho \right) ,  \label{5}
\end{equation}%
where the vacuum state $\left \vert 0\tilde{0}\right \rangle $ is
annihilated by either $a$ or\ $\tilde{a}.$ The parameter
\begin{equation}
\tanh \theta =\exp \left( -\frac{\hbar \omega }{2kT}\right) ,  \label{7}
\end{equation}%
is determined by comparing the Bose-Einstein distribution
\begin{equation}
\text{Tr}\left( \rho _{c}a^{\dagger }a\right) =\left[ e^{\omega \hbar /kT}-1%
\right] ^{-1}\equiv \bar{n},  \label{8}
\end{equation}%
with the expectation value of the photon number operator in $\left \vert
0(\beta )\right \rangle $%
\begin{equation}
\left \langle 0(\beta )\right \vert a^{\dagger }a\left \vert 0(\beta )\right
\rangle =\sinh ^{2}\theta .  \label{9}
\end{equation}%
The reason we tackle with $\left \vert 0\left( \beta \right) \right \rangle
\left \langle 0\left( \beta \right) \right \vert $ lies in that partial
tracing over its tilde-mode will lead to $\rho _{c}$ (this will be proved in
Sec. 2), i.e.%
\begin{equation}
\text{\~{T}r}\left[ \left \vert 0\left( \beta \right) \right \rangle \left
\langle 0\left( \beta \right) \right \vert \right] =\rho _{c}=\left( 1-e^{-%
\frac{\hbar \omega }{kT}}\right) e^{-\frac{\hbar \omega }{kT}a^{\dagger }a},
\label{10}
\end{equation}%
then when we take $\left \vert 0(\beta )\right \rangle \left \langle 0(\beta
)\right \vert $ as $\rho _{0}$ and substitute it into Eq. (\ref{2})
\begin{equation}
\rho _{c}\left( t\right) =\sum_{n=0}^{\infty }\frac{V^{n}}{n!}e^{-\kappa
ta^{\dagger }a}a^{n}\left \vert 0\left( \beta \right) \right \rangle \left
\langle 0\left( \beta \right) \right \vert a^{\dag n}e^{-\kappa ta^{\dagger
}a},  \label{11}
\end{equation}%
we will have
\begin{eqnarray}
\text{\~{T}r}\left[ \rho _{c}\left( t\right) \right] &=&\sum_{n=0}^{\infty }%
\frac{V^{n}}{n!}e^{-\kappa ta^{\dagger }a}a^{n}\left[ t\tilde{r}\left \vert
0\left( \beta \right) \right \rangle \left \langle 0\left( \beta \right)
\right \vert \right] a^{\dag n}e^{-\kappa ta^{\dagger }a}  \label{12} \\
&=&\sum_{n=0}^{\infty }\frac{T^{\prime n}}{n!}e^{-\kappa ta^{\dagger
}a}a^{n}\rho _{c}a^{\dag n}e^{-\kappa ta^{\dagger }a},  \notag
\end{eqnarray}%
thus \~{T}r$\left[ \rho _{c}\left( t\right) \right] $ will present the
correct dissipation evolution law of the chaotic field. Noting that $\rho
_{c}\left( t\right) $ itself includes not only the information of the
system, but also of the reservoir, and we can also see how the reservoir
evolves accompanying the system's dissipation. Moreover, since the
temperature effect is manifest through the structure of $\rho _{c}$, we also
investigate how system's dissipation accompanies the temperature variation.

\section{Partial tracing over the tilde-mode of $\left \vert 0\left( \protect%
\beta \right) \right \rangle \left \langle 0\left( \protect \beta \right)
\right \vert $}

Using the coherent state representation of the tilde-mode%
\begin{equation}
\int \frac{d^{2}z}{\pi }\left \vert \tilde{z}\right \rangle \left \langle
\tilde{z}\right \vert =1,\text{ \  \  \ }\tilde{a}\left \vert \tilde{z}\right
\rangle =z\left \vert \tilde{z}\right \rangle ,  \label{13}
\end{equation}%
where
\begin{equation}
\left \vert \tilde{z}\right \rangle =\exp \left[ -\frac{|z|^{2}}{2}+z\tilde{a%
}^{\dagger }\right] \left \vert \tilde{0}\right \rangle ,  \label{14}
\end{equation}%
we have%
\begin{eqnarray}
\text{\~{T}r}\left[ \left \vert 0\left( \beta \right) \right \rangle \left
\langle 0\left( \beta \right) \right \vert \right] &=&\text{\~{T}r}\left[
\int \frac{d^{2}z}{\pi }\left \vert \tilde{z}\right \rangle \left \langle
\tilde{z}\right \vert \left. 0\left( \beta \right) \right \rangle \left
\langle 0\left( \beta \right) \right \vert \right]  \notag \\
&=&\sec h^{2}\theta \int \frac{d^{2}z}{\pi }\left \langle \tilde{z}\right
\vert e^{a^{\dag }z^{\ast }\tanh \theta }\left \vert 0\tilde{0}\right
\rangle \left \langle 0\tilde{0}\right \vert e^{az\tanh \theta }\left \vert
\tilde{z}\right \rangle .  \label{15}
\end{eqnarray}%
Then using $\left \langle \tilde{z}\right \vert \left. \tilde{0}%
\right
\rangle =\exp \left( -\left \vert z\right \vert ^{2}/2\right) $, the
normal ordering of $\left \vert 0\right \rangle \left \langle 0\right \vert $%
\begin{equation}
\left \vert 0\right \rangle \left \langle 0\right \vert =\colon
e^{-a^{\dagger }a}\colon ,  \label{16}
\end{equation}%
and%
\begin{equation*}
e^{\lambda a^{\dagger }a}=\colon \exp \left[ \left( e^{\lambda }-1\right)
a^{\dagger }a\right] \colon
\end{equation*}%
we have%
\begin{equation}
\text{\~{T}r}\left[ \left \vert 0\left( \beta \right) \right \rangle \left
\langle 0\left( \beta \right) \right \vert \right] =\sec h^{2}\theta \int
\frac{d^{2}z}{\pi }\colon e^{-\left \vert z\right \vert ^{2}+a^{\dag
}z^{\ast }\tanh \theta +az\tanh \theta -a^{\dag }a}\colon =\sec h^{2}\theta
\colon e^{a^{\dag }a\left( \tanh ^{2}\theta -1\right) }\colon  \label{17}
\end{equation}%
and noting $\tanh \theta =\exp \left( -\tfrac{\hbar \omega }{2kt}\right) $
\begin{equation}
\text{\~{T}r}\left[ \left \vert 0\left( \beta \right) \right \rangle \left
\langle 0\left( \beta \right) \right \vert \right] =\left[ 1-\exp \left( -%
\tfrac{\hbar \omega }{2kt}\right) \right] \exp \left( -\tfrac{\hbar \omega }{%
2kt}a^{\dag }a\right) =\rho _{c}.  \label{18}
\end{equation}
Note that the partial trace over\ mode $a$ for $\left \vert 0\left( \beta
\right) \right \rangle \left \langle 0\left( \beta \right) \right \vert $ is
Tr$\left[ \left \vert 0\left( \beta \right) \right \rangle \left \langle
0\left( \beta \right) \right \vert \right] =\left( 1-e^{-\frac{\hbar \omega
}{kT}}\right) e^{-\frac{\hbar \omega }{kT}\tilde{a}^{\dagger }\tilde{a}},$
since in Eq. (\ref{6}) $\left \vert 0\left( \beta \right) \right \rangle $
is symmetric with respect to $\tilde{a}^{\dagger }$ and $a^{\dagger }.$

Remarkably, the thermo vacuum state can be rewritten as
\begin{equation}
\left \vert 0(\beta )\right \rangle =\sec h\theta \exp \left[ a^{\dagger }%
\tilde{a}^{\dagger }\tanh \theta \right] \left \vert 0\tilde{0}\right
\rangle =S\left( \theta \right) \left \vert 0\tilde{0}\right \rangle
\label{19}
\end{equation}%
where%
\begin{equation}
S\left( \theta \right) =\exp \left[ \theta \left( a^{\dagger }\tilde{a}%
^{\dagger }-a\tilde{a}\right) \right]  \label{20}
\end{equation}%
is in form like a two-mode squeezing operator \cite{Wang_cpl_2010}, so $%
S\left( \theta \right) $ is named thermo squeezing operator which squeezes
the vacuum state $\left \vert 0\tilde{0}\right \rangle $ at zero-temperature
to the thermo vacuum state $\left \vert 0(\beta )\right \rangle $ at finite
temperature $T$. Because a two-mode squeezed state is an entangled state, so
$\left \vert 0(\beta )\right \rangle $ can be considered an entangled state
in which the syetem's mode $a^{\dagger }$ entangles with the tilde mode $%
\tilde{a}^{\dagger }$. Since they are entangled, the dissipation of system
will affect its enviroment, as we shall show in the next section.

\section{Evolution of $\left \vert 0\left( \protect \beta \right)
\right
\rangle \left \langle 0\left( \protect \beta \right) \right \vert $
in dissipation channel}

Using
\begin{equation}
a^{n}\left \vert 0\left( \beta \right) \right \rangle =a^{n-1}\sec \text{h}%
\theta \left[ a^{n},e^{a^{\dagger }\tilde{a}^{\dagger }\tanh \theta }\right]
\left \vert 0\tilde{0}\right \rangle =\left( \tilde{a}^{\dagger }\tanh
\theta \right) ^{n}\left \vert 0\left( \beta \right) \right \rangle ,
\label{21}
\end{equation}%
we obtain%
\begin{align}
\rho _{c}\left( t\right) & =\sum_{n=0}^{\infty }\frac{V^{n}\tanh ^{2n}\theta
}{n!}e^{-\kappa ta^{\dagger }a}\tilde{a}^{\dagger n}\left \vert 0\left(
\beta \right) \right \rangle \left \langle 0\left( \beta \right) \right
\vert \tilde{a}^{n}e^{-\kappa ta^{\dagger }a}  \label{22} \\
& =\sec \text{h}^{2}\theta \sum_{n=0}^{\infty }\frac{V^{n}\tanh ^{2n}\theta
}{n!}e^{-\kappa ta^{\dagger }a}\tilde{a}^{\dagger n}e^{a^{\dagger }\tilde{a}%
^{\dagger }\tanh \theta }\left \vert 0\tilde{0}\right \rangle \left \langle 0%
\tilde{0}\right \vert e^{a\tilde{a}\tanh \theta }\tilde{a}^{n}e^{-\kappa
ta^{\dagger }a}
\end{align}%
Then using
\begin{equation}
e^{-\kappa ta^{\dagger }a}a^{\dagger }e^{\kappa ta^{\dagger }a}=e^{-\kappa
t}a^{\dagger }  \label{23}
\end{equation}%
we obtain%
\begin{equation}
\rho _{c}\left( t\right) =\sec \text{h}^{2}\theta \sum_{n=0}^{\infty }\frac{%
V^{n}\tanh ^{2n}\theta }{n!}\tilde{a}^{\dagger n}e^{e^{-\kappa t}a^{\dagger }%
\tilde{a}^{\dagger }\tanh \theta }\left \vert 0\tilde{0}\right \rangle \left
\langle 0\tilde{0}\right \vert e^{e^{-\kappa t}a\tilde{a}\tanh \theta }%
\tilde{a}^{n}.  \label{24}
\end{equation}%
Further, by using the normal product form
\begin{equation}
\left \vert 0\tilde{0}\right \rangle \left \langle 0\tilde{0}\right \vert
=\colon e^{-a^{\dagger }a-\tilde{a}^{\dagger }\tilde{a}}\colon ,  \label{25}
\end{equation}%
we can make summation in Eq. (\ref{24}) and derive the compact form of $\rho
\left( t\right) ,$

\begin{align}
\rho _{c}\left( t\right) & =\sec h^{2}\theta \colon \sum_{n=0}^{\infty }%
\frac{V^{n}\tanh ^{2n}\theta }{n!}\tilde{a}^{\dagger n}\tilde{a}%
^{n}e^{e^{-\kappa t}a^{\dagger }\tilde{a}^{\dagger }\tanh \theta
}e^{e^{-\kappa t}a\tilde{a}\tanh \theta -a^{\dagger }a-\tilde{a}^{\dagger }%
\tilde{a}}\colon  \notag \\
& =\sec \text{h}^{2}\theta \colon \exp \{ \tilde{a}^{\dagger }\tilde{a}%
\left( 1-e^{-2\kappa t}\right) \tanh ^{2}\theta +e^{-\kappa t}\tanh \theta
\left( a^{\dagger }\tilde{a}^{\dagger }+a\tilde{a}\right) -a^{\dagger }a-%
\tilde{a}^{\dagger }\tilde{a}\} \colon  \label{26} \\
& =\sec \text{h}^{2}\theta e^{e^{-\kappa t}a^{\dagger }\tilde{a}^{\dagger
}\tanh \theta }\left \vert 0\right \rangle \left \langle 0\right \vert \exp
\{ \tilde{a}^{\dagger }\tilde{a}\ln \left[ \left( 1-e^{-2\kappa t}\right)
\tanh ^{2}\theta \right] \}e^{e^{-\kappa t}\tanh \theta a\tilde{a}}  \notag
\end{align}%
where $\exp \{ \tilde{a}^{\dagger }\tilde{a}\ln \left[ \left( 1-e^{-2\kappa
t}\right) \tanh ^{2}\theta \right] \}$ indicates that the reservoir is in a
chaotic field of the tilde mode, no more in $\left \vert \tilde{0}%
\right
\rangle \left \langle \tilde{0}\right \vert ,$ this is because the
system mode and the reservior mode are entangled, the disspation of system
certainly affects the reservoir. From (\ref{26}) we can realize how a pure
thermo vacuum state evolves into the mixed state during the dissipation
process, that is, not only the squeezing parameter $\tanh \theta \rightarrow
e^{-\kappa t}\tanh \theta ,$ but also $\left \vert 0\tilde{0}\right \rangle
\left \langle 0\tilde{0}\right \vert $ evolves into $\left \vert
0\right
\rangle \left \langle 0\right \vert \exp \{ \tilde{a}^{\dagger }%
\tilde{a}\ln \left[ \left( 1-e^{-2\kappa t}\right) \tanh ^{2}\theta \right]
, $ i.e., in the process the thermo squeezing effect decreases while the
reservioir-mode vacuum becomes chaotic.

\section{Partial traceing over the tilde-mode of $\protect \rho _{c}\left(
t\right) $}

Now we perform partial trace over the tilde-mode of $\rho _{c}\left(
t\right) ,$ using (\ref{13}), (\ref{16}-\ref{18}) and (\ref{26}) we have%
\begin{eqnarray*}
\text{\~{T}r}\left[ \rho _{c}\left( t\right) \right]  &=&\sec h^{2}\theta
\text{\~{T}r}\left[ \frac{d^{2}z}{\pi }\left \vert \tilde{z}\right \rangle
\left \langle \tilde{z}\right \vert \exp \left[ e^{-kt}a^{\dag }\tilde{a}%
^{\dag }\tanh \theta \right] \left \vert 0\right \rangle \left \langle
0\right \vert \exp \left \{ \tilde{a}^{\dagger }\tilde{a}\ln \left[ \left(
1-e^{-2\kappa t}\right) \tanh ^{2}\theta \right] \right \} \exp \left[ a%
\tilde{a}e^{-kt}\tanh \theta \right] \right]  \\
&=&\sec h^{2}\theta \int \frac{d^{2}z}{\pi }\colon \exp \left \{ \left \vert
z\right \vert ^{2}\left[ \left( 1-e^{-2kt}\right) \tanh ^{2}\theta -1\right]
+e^{-kt}\left( a^{\dagger }z^{\ast }+az\right) \tanh \theta -a^{\dagger
}a\right \} \colon  \\
&=&\frac{1}{1+e^{-2\kappa t}\sinh ^{2}\theta }\exp \left[ a^{\dagger }a\ln
\frac{e^{-2\kappa t}\tanh ^{2}\theta }{1-\left( 1-e^{-2\kappa t}\right)
\tanh ^{2}\theta }\right] .
\end{eqnarray*}%
When $t=0,$ it becomes the original chaotic state. By identifying%
\begin{equation}
\frac{e^{-\kappa t}\tanh \theta }{\sqrt{1-\left( 1-e^{-2\kappa t}\right)
\tanh ^{2}\theta }}=\tanh \theta ^{\prime },  \label{28}
\end{equation}%
then%
\begin{equation}
\frac{1}{1+e^{-2\kappa t}\sinh ^{2}\theta }=\sec h^{2}\theta ^{\prime }
\label{29}
\end{equation}%
and we can express%
\begin{equation}
\text{\~{T}r}\left[ \rho _{c}\left( t\right) \right] =\sec \text{h}%
^{2}\theta ^{\prime }\exp [a^{\dagger }a\ln \tanh ^{2}\theta ^{\prime }],
\label{30}
\end{equation}%
which explains that the system in still in a chaotic state but with new
parameter $\theta ^{\prime }.$ In similar to Eq. (\ref{7}), by identifying%
\begin{equation}
\ln \frac{e^{-2\kappa t}\tanh ^{2}\theta }{1-\left( 1-e^{-2\kappa t}\right)
\tanh ^{2}\theta }=-\frac{\hbar \omega }{kT^{\prime }},  \label{31}
\end{equation}%
we see that system is now at the temperature%
\begin{equation}
T^{\prime }=-\frac{\hbar \omega }{k\ln \frac{e^{-2\kappa t}\tanh ^{2}\theta
}{1-\left( 1-e^{-2\kappa t}\right) \tanh ^{2}\theta }}.  \label{32}
\end{equation}%
Due to
\begin{equation}
1-\left( 1-e^{-2\kappa t}\right) \tanh ^{2}\theta >e^{-2\kappa t},
\label{33}
\end{equation}%
so%
\begin{equation}
\ln \frac{e^{-2\kappa t}\tanh ^{2}\theta }{1-\left( 1-e^{-2\kappa t}\right)
\tanh ^{2}\theta }<0,\text{ \  \  \  \ }T^{\prime }>0.  \label{34}
\end{equation}%
Moreover, since%
\begin{equation}
\tanh ^{2}\theta >\frac{e^{-2\kappa t}\tanh ^{2}\theta }{1-\left(
1-e^{-2\kappa t}\right) \tanh ^{2}\theta },  \label{35}
\end{equation}%
\begin{equation}
-\frac{1}{\ln \tanh ^{2}\theta }>\frac{-1}{\ln \frac{e^{-2\kappa t}\tanh
^{2}\theta }{1-\left( 1-e^{-2\kappa t}\right) \tanh ^{2}\theta }},
\label{36}
\end{equation}%
and in reference to Eq. (\ref{7}), $\ln \tanh ^{2}\theta =-\frac{\hbar
\omega }{kT},$ $T=-\frac{\hbar \omega }{k\ln \tanh ^{2}\theta },$ we see%
\begin{equation}
T>T^{\prime }  \label{37}
\end{equation}%
which states that during the damping process the system's temperature
decreases, the rate of decreasing can be controlled by adjusting the damping
rate $\kappa .$

In summary, we have examined temperature cooling of optical chaotic light in
a quantum dissipation channel with the damping parameter $\kappa .$The way
we do it is by introducing its thermal vacuum state which can expose
entangling effect between the system and the reservoir. The temperature
cooling formula (\ref{32}) is derived, which depends on the parameter $%
\kappa ,$ by adjusting $\kappa $ one can control temperature.

\end{document}